\documentclass[]{spie}
\pdfoutput=1
\usepackage{amsmath,amsfonts,amssymb}
\usepackage{graphicx}
\usepackage[colorlinks=true, allcolors=blue]{hyperref}
\usepackage{listings}
\usepackage{xcolor}

\newcommand{\aj}{AJ}
\newcommand{\apj}{ApJ}
\newcommand{\apjs}{ApJS}
\newcommand{\procspie}{Proc.\ SPIE}
\newcommand{\pasj}{PASJ}

\newcommand\YAMLcolonstyle{\color{red}\mdseries}
\newcommand\YAMLkeystyle{\color{black}\bfseries}
\newcommand\YAMLvaluestyle{\color{blue}\mdseries}

\lstdefinelanguage{YAML}{
    keywords={true,false,null,y,n},
    keywordstyle=\color{darkgray}\bfseries,
    basicstyle=\YAMLkeystyle,                                 % assuming a key comes first
    sensitive=false,
    comment=[l]{\#},
    morecomment=[s]{/*}{*/},
    commentstyle=\color{purple}\ttfamily,
    stringstyle=\YAMLvaluestyle\ttfamily,
    moredelim=[l][\color{orange}]{\&},
    moredelim=[l][\color{magenta}]{*},
    moredelim=**[il][\YAMLcolonstyle{:}\YAMLvaluestyle]{:},   % switch to value style at :
    morestring=[b]',
    morestring=[b]",
    literate =    {---}{{\ProcessThreeDashes}}3
    {>}{{\textcolor{red}\textgreater}}1
    {|}{{\textcolor{red}\textbar}}1
    {\ -\ }{{\mdseries\ -\ }}3,
}

\title{The Vera C. Rubin Observatory Data Butler and Pipeline Execution System}

\author[1]{Tim~Jenness}
\affil[1]{Rubin Observatory Project Office, 950 N.\ Cherry Ave., Tucson, AZ  85719, USA}
\author[2]{James~F.~Bosch}
\affil[2]{Department of Astrophysical Sciences, Princeton University, Princeton, NJ 08544, USA}

\author[3]{Andrei~Salnikov}
\affil[3]{SLAC National Accelerator Laboratory,  2575 Sand Hill Rd., Menlo Park, CA 94025, USA}
\author[2]{Nate~B.~Lust}
\author[3]{Nathan~M.~Pease}
\author[4]{Michelle~Gower}
\affil[4]{NCSA, University of Illinois at Urbana-Champaign, 1205 W.\ Clark St., Urbana, IL 61801, USA}
\author[4]{Mikolaj~Kowalik}
\author[5]{Gregory~P.~Dubois-Felsmann}
\affil[5]{IPAC, California Institute of Technology, MS 100-22, Pasadena, CA 91125, USA}
\author[3]{Fritz~Mueller}
\author[2]{Pim~Schellart}

\begin{document}
\maketitle

\begin{abstract}
    The Rubin Observatory's Data Butler is designed to allow data file location and file formats to be abstracted away from the people writing the science pipeline algorithms.
    The Butler works in conjunction with the workflow graph builder to allow pipelines to be constructed from the algorithmic tasks.
    These pipelines can be executed at scale using object stores and multi-node clusters, or on a laptop using a local file system.
    The Butler and pipeline system are now in daily use during Rubin construction and early operations.
\end{abstract}


\section{Introduction}

The Vera C.\ Rubin Observatory's Legacy Survey of Space and Time (LSST) \cite{2019ApJ...873..111I} will image the entire southern sky every three days and consist of tens of petabytes of raw image data and associated calibration data.
All these files must be tracked, along with the intermediate datasets and output products from pipeline processing, and depending on where the processing occurs the files will be stored on either POSIX file systems or object stores.
The LSST Data Management System (DMS)\cite{2017ASPC..512..279J} is responsible for transferring raw files off the mountain and storing them at the US Data Facility (USDF).
The datasets are then processed by pipelines\cite{2019ASPC..523..521B,2018PASJ...70S...5B} that can be run at different data centers and the results integrated into a unified data release.
This paper will discuss the part of the DMS that abstracts data access from the pipeline algorithms, builds the execution workflow graphs, and allows the processing jobs to be run in large batch processing systems.

\section{The Data Butler}

The Data Butler (hereafter the ``Butler'') is the system that abstracts the data access details from the pipeline developers.
The top requirements for the Butler are:

\begin{itemize}
\item The pipeline writer should not have to know where the data are being read from or written to or what file formats are being used, or even if a file is involved at all.
\item The pipeline writer only has to deal with Python objects.
\item Users should be able to locate the relevant data using common astronomical concepts such as observation identifier, physical filter, patch of sky, or telescope.
\end{itemize}

The current implementation of the Butler is colloquially known as the ``Generation 3'' Butler since it is the third implementation during the evolution of the LSST DMS, a project that began nearly 20 years ago.\cite{2006SPIE.6274E..0PK}
The ``Generation 2'' Butler was used for many years but certain limitations in the design and implementation, especially in terms of its relationship with the evolving demands of the pipeline design and need for more flexibility, led to it being completely replaced with a brand new system that shares some concepts but no code with previous versions.

The two main components of the Butler are the Registry and the Datastore.
Registry organizes datasets conceptually and associates them with astronomical concepts as stated above, but it has no idea where or how datasets are persisted.
Datastore is responsible for serializing a Python object to a storage system and reading data back from that storage system and recreating the Python object.
There is then a thin layer of code (the responsibility of the \texttt{Butler} Python class itself) providing a unified interface to the user that coordinates the Datastore and Registry interaction to ensure consistency.

\subsection{Defining a Dataset}

A cornerstone of the Butler design is that any dataset can be found by specifying a coordinate in its dimensional space.
These dimensions are generally quantities that can be understood by an astronomer as being relevant for the particular dataset.
For example, raw data can be addressed using dimensions \texttt{instrument}, \texttt{detector}, and \texttt{exposure}.
A specific dataset must then be specified using a coordinate within that dimensional space, for example: \texttt{\{instrument="LSSTCam", detector=1, exposure=2024050100023\}} will uniquely represent the second detector of the 23rd LSSTCam observation taken on 2024-05-01.\footnote{Embedding the date in the integer exposure ID is an LSSTCam convention and is not required. Any integer can be used here so long as it uniquely identifies a specific observation.}

In order to determine which dimensions are relevant for a specific dataset, Butler has the concept of a ``dataset type''.
A dataset type gives a name to the combination of the list of relevant dimensions, an abstraction around the Python type it corresponds to, and whether or not it is a calibration dataset.
A calibration dataset is a dataset that is associated with a data coordinate (for example an instrument and detector) but can be used for calibrating datasets, such as those used for instrument signature removal, that is additionally associated with a temporal validity range.
The name of the dataset type itself does not matter, but it is important that the names are used consistently because they are critical when constructing pipelines.
Raw data are commonly given the dataset type \texttt{raw} and the LSST science pipelines use dataset type names such as \texttt{diffim} (difference image), \texttt{defects} (pixel defect masks), and \texttt{bias} (processed biases), with the intent being for the name to be descriptive enough for a science user to have a good idea as to what type of data it represents.

The abstraction around the Python type is called a ``storage class''.
The Butler's main job is to store Python objects somewhere and to retrieve them again.
We use storage classes to map a name to a Python type and allow additional configuration to be associated with the Python type such as whether Butler can treat the type as a composite and what parameters can be included in the retrieval to modify the returned object.
Composites can be important in that some pipelines may, for example, only need to take a WCS or a pixel mask as input and do not want the entire calibrated exposure to be loaded.
A storage class can declare the components it allows and then declare the associated storage classes of each component.

The data model used by Butler is currently designed specifically for astronomical imaging data.
The model itself is specified in a YAML text file and is therefore easily extended or modified to suit specific needs.
There is a tension between trying to develop one unified model that will support all astronomy data versus an approach where every observatory specifies their own dimension universe for every instrument.
A unified model is needed in scenarios where datasets from different instruments should be processed in the same pipeline.
There is, though, no problem with multiple distinct Butler repositories being configured with their own dimension universes when the data are very different and then requiring the user to choose the relevant Butler repository.
Currently there are two projects experimenting with the Butler using non-imaging survey data, and both Subaru's Prime Focus Spectrograph (PFS) team\cite{2020SPIE11447E..7VW} and NASA's SPHEREx team\cite{2020SPIE11443E..0IC} have decided to alter the default dimension universe data model.

\subsection{Collections}

A dataset can be specified by its data coordinate and dataset type, but the Butler needs one final piece of information to uniquely locate the desired dataset.
In order to allow different pipeline execution runs to store their outputs without over-writing datasets from previous runs, each dataset must be associated with a single \texttt{RUN} collection.
Every dataset tracked in a Butler Registry can be uniquely located by knowing the run collection, data coordinate, and dataset type.

Whilst every dataset must be stored in a \texttt{RUN} collection there are also other types of collections that can be used:

\begin{itemize}
\item A \texttt{TAGGED} collection is a collection of arbitrary datasets.
\item A \texttt{CHAINED} collection is a collection of other collections and those other collections can be of any type (chains of chains are allowed).
      A \texttt{CHAINED} collection does not itself contain datasets.
      The order of collections in a chain is important and determines how datasets are found.
\item A \texttt{CALIBRATION} collection is a special type of collection where a calibration dataset (one whose dataset type is marked as a calibration) can be associated with a validity timespan.
\end{itemize}

All these collections are important when building and executing pipelines.

\subsection{The Registry}

Registry can be thought of as a database that can be queried to find out what datasets are available.
We define Registry to have an abstract interface to allow different implementations to co-exist, although we do assume that a SQL-like syntax is supported when constraining queries (effectively a \texttt{WHERE} clause), even if that does not result in SQL being executed in the specific Registry implementation.

The primary Registry implementation uses SQLAlchemy \cite{myers2015essential}\footnote{\url{https://www.sqlalchemy.org}} to talk to PostgreSQL\footnote{\url{https://www.postgresql.org}} or SQLite\footnote{\url{https://www.sqlite.org/index.html}} backend databases.
We do not make use of the Object Relational Mapping interface, and instead use the low-level SQLAlchemy ``Core'' interface to create and interact with tables.

The Registry stores all the dimension information, collection and dataset type definitions, and the associations between these and the actual datasets.
All dimension values must be pre-defined before a dataset can be stored with that value, and the dimension records have additional metadata that can be used to constrain dataset queries or to provide more detail as to what a dimension really represents.
For example, in the default schema, which is currently used by LSST cameras, Hyper Suprime-Cam\cite{2018PASJ...70S...1M} and DECam\cite{2015AJ....150..150F}, the \texttt{exposure} dimension contains metadata including the time of the exposure, the elevation and azimuth, the observation type, and the exposure time, whereas the \texttt{detector} dimension contains metadata including the full name, the raft name, and the detector role.
The tables that hold this dimension metadata exist independently of the datasets they correspond to; for example, while the metadata associated with the \texttt{exposure} dimension metadata is usually derived from the header of a \texttt{raw} dataset, that \texttt{exposure} is also associated with the dataset types that result from processing done on that exposure.
The dimension tables -- which often also include foreign key relationships, such as the \texttt{physical\_filter} associated with an \texttt{exposure} -- thus form a data model ``skeleton'' of sorts for the actual datasets, which have no relationships of their own.
In some cases those dimension relationships are spatial (some dimensions are associated with regions on the sky) or temporal (associated with time spans).

These dimension relationships allow the Registry to provide a rich query system, based on a custom expression language based on SQL boolean expressions.
These are parsed and translated by Registry into actual SQL queries, and we delegate optimization and execution of these queries almost entirely to the underlying database.
But in contrast to raw SQL the user essentially has to only provide the Registry the equivalent of the \texttt{WHERE} clause, because the \texttt{SELECT} clause is based on what dataset types or dimensions are being requested, and the multi-join \texttt{FROM} clause can be worked out internally according to the known dimension relationships, despite those relationships being encoded in YAML configuration rather than code.
The query system also isolates the actual database schema from the user, allowing it to be changed for performance reasons without breaking user code.

\subsection{The Datastore}

Datastores are responsible for serializing a Python object to a storage system and reading the data back from the storage system and returning a Python object.
On \texttt{put}, Datastore is given a reference to the dataset (encapsulated in the Python class \texttt{DatasetRef}) by Registry (this is the data coordinate, the dataset type, the run collection, and the unique identifier, a UUID,\cite{rfc4122uuid} assigned by Registry) and the Python object.
To retrieve a dataset the Datastore is passed in the dataset reference.
By design the only connection between Registry and Datastore is the dataset reference, although in some implementations the Datastore can make use of the shared PostgreSQL database to store relevant information.

The Butler user does not have to know how data are serialized or where it is coming from, indeed there is no requirement for files to be involved at all.
In the current system there are three Datastores defined: \texttt{FileDatastore} serializes to files; \texttt{InMemoryDatastore} stores Python objects in an in-memory cache; and \texttt{ChainedDatastore} is a Datastore consisting of other Datastores.
All Datastores support configuration-based constraints that can be used to decide whether a specific dataset type or storage class should be accepted or rejected by the Datastore.

\subsubsection{File Datastore}

The \texttt{FileDatastore} is the most fundamental of the Datastore implementations in that it is responsible for serializing datasets to files and reading them back in again.
A file Datastore is defined by a URI pointing to the area where files are under the control of the Datastore.
A URI is used rather than a file path to allow file I/O itself to be abstracted to allow use of remote storage or POSIX file systems.
To achieve this abstraction we use a unified URI handler class, \texttt{ResourcePath} from the \texttt{lsst-resources} package\footnote{\url{https://github.com/lsst/resources}}, which currently supports S3 storage, Google Cloud Storage, WebDAV, and POSIX, as well as new storage systems which can easily be added.

Within the file Datastore the class that is responsible for creating the file and reading the file is called the \texttt{Formatter}.
The Datastore configuration system includes a look up table that matches the dataset type or storage class of the relevant dataset to a corresponding formatter implementation.
This formatter is then given the Python object and the destination location and told to write the file.
Similarly, when a dataset is requested, the formatter is looked up in the internal Datastore database and the formatter is told to read the file.
We do not look up the formatter from configuration when reading since it is possible that the configuration may have changed since the file was stored.

Whereas the Registry only deals with datasets, the Datastore understands composites.
When a dataset is associated with a composite storage class, the Datastore can be configured to disassemble the dataset and write the contents as distinct files.
For example, an astronomical image could be disassembled into the pixel image data, a variance plane, a mask, and a header before being written out as four files.
If a user requests just the header component, that file can be read without needing to look at the other files, and if the full dataset is requested all components will be read and the composite will be reassembled using the helper class declared in the storage class definition.
Of course, if a user requests a component when the dataset was not disassembled, this will still work, although it may require the entire file be downloaded from a remote object store in order to extract a small subset.

When the datasets are stored on a slow network disk or a remote object store, it is inefficient to continually re-read the files if the pipeline algorithm is reading subsets or if the user is requesting individual components.
The file Datastore overcomes this by implementing a caching system where the file is stored in a more local cache directory on first retrieval.
The cache can expire files based on number of files, total cache size, or number of datasets (noting that a disassembled composite will consist of multiple files).
In some situations this can significantly improve performance with remote Datastores.

\subsubsection{In-Memory Datastore}

This is an in-memory caching Datastore.
Since it caches Python objects in a single process its main purpose is to support intermediate products that are to be passed from one task to another without needing to include the overhead for serialization.
In most scenarios an in-memory Datastore is combined with another Datastore in a chain.
Care must be taken when returning datasets from the in-memory Datastore since Python can not prevent the caller from modifying the object, which would cause confusion if the object is retrieved a second time.
Options being considered are to always deep copy the object on return, or consider allowing the Datastore to be configured to remove the object from the cache when retrieved.

\subsubsection{Chained Datastore}

The chained Datastore implementation does not itself store any datasets.
It consists of multiple Datastores, each with its own configuration and allowed to be of any type.
When writing a dataset the dataset is presented to each Datastore in turn and everything is okay so long as one Datastore accepts the dataset.
When reading a dataset each Datastore is asked for the dataset in turn and the Python object is returned from the first matching Datastore.
Combined with per-Datastore dataset type constraints, this can allow some datasets to be stored in an in-memory Datastore and file Datastore but allow the in-memory dataset to be retrieved efficiently, whilst other datasets are only stored in the file Datastore.
Alternatively a chain of two file Datastores can be used to allow one Datastore to be read-only (for example a Datastore containing a validated self-contained data release) and a second Datastore to accept derived products from users.

\subsubsection{Other Datastores}

The Datastore interface does not require datasets to be persisted as files.
For example, we are considering storing pipeline metrics directly into our metrics database \cite{SQR-019,DMTN-203}.
The Butler user would not know whether a metric was being stored as a JSON file in a file Datastore or stored directly into a metrics database (or even stored in both places).

\subsection{Client/Server Butler}

The Butler is a Python library that, by default, is set up to use SQLAlchemy to talk to a SQL Registry, and will use AWS or Google Cloud credentials to talk to object stores.
This is not a convenient interface if Python is not being used, or if the authorization required to access the database or object store directly is not available to the people trying to access the Butler.
The latter situation is likely to be the situation in the Rubin Science Platform in the cloud where users will be logged in with Rubin accounts but will not be issued cloud credentials.\cite{DMTN-182,2021arXiv211115030O}.

To solve this problem we are working on a \texttt{https} client/server Butler\cite{DMTN-176}.
The client user will present their authentication token and the server will then determine if that user is authorized to retrieve the requested dataset.
If they are allowed, they will be returned a signed URL that their client can then use to retrieve the dataset.

\subsection{Command Line Tooling}

\begin{figure}
    \begin{small}
    \begin{verbatim}
        $ butler --help
        Options:
          --log-level LEVEL|COMPONENT=LEVEL ...
                                          The logging level. Without an explicit
                                          logger name, will only affect the default
                                          root loggers (lsst). To modify the root
                                          logger use '.=LEVEL'. Supported levels are [
                                          CRITICAL|ERROR|WARNING|INFO|VERBOSE|DEBUG|TR
                                          ACE]
          --long-log                      Make log messages appear in long format.
          --log-file FILE ...             File(s) to write log messages. If the path
                                          ends with '.json' then JSON log records will
                                          be written, else formatted text log records
                                          will be written. This file can exist and
                                          records will be appended.
          --log-tty / --no-log-tty        Log to terminal (default). If false logging
                                          to terminal is disabled.
          --log-label TEXT ...            Keyword=value pairs to add to MDC of log
                                          records.
          --progress / --no-progress      Show a progress bar for slow operations when
                                          possible.
          -h, --help                      Show this message and exit.

        Commands:
          associate                   Add existing datasets to a tagged collection.
          certify-calibrations        Certify calibrations in a repository.
          collection-chain            Define a collection chain.
          config-dump                 Dump butler config to stdout.
          config-validate             Validate the configuration files.
          convert                     Convert a gen2 repo to gen3.
          create                      Create an empty Gen3 Butler repository.
          define-visits               Define visits from exposures.
          export-calibs               Export calibrations from the butler for later import.
          import                      Import data into a butler repository.
          ingest-files                Ingest files from table file.
          ingest-photodiode           Ingest photodiode data.
          ingest-raws                 Ingest raw frames.
          make-discrete-skymap        Define a discrete skymap from calibrated exposures.
          prune-datasets              Remove datasets.
          query-collections           Search for collections.
          query-data-ids              List the data IDs in a repository.
          query-dataset-types         Get the dataset types in a repository.
          query-datasets              List the datasets in a repository.
          query-dimension-records     Query for dimension information.
          register-dataset-type       Register a new dataset type with this...
          register-dcr-subfilters     Add subfilters for chaotic modeling.
          register-instrument         Add an instrument definition to the repository
          register-skymap             Make a SkyMap and add it to a repository.
          remove-collections          Remove one or more non-RUN collections.
          remove-dataset-type         Remove a dataset type definition from a repository.
          remove-runs                 Remove one or more RUN collections.
          retrieve-artifacts          Retrieve file artifacts from a Butler.
          transfer-datasets           Transfer datasets from one butler to another.
          write-curated-calibrations  Add an instrument's curated calibrations.
    \end{verbatim}
    \end{small}
    \caption{The \texttt{butler} command line options and subcommands.}
    \label{fig:butler-cli}
\end{figure}

There are many common operations that should be achievable without needing to write any Python code.
To serve those use cases we provide command-line tooling based on the \texttt{click} Python package.\footnote{\url{https://click.palletsprojects.com}}
Examples of the current subcommands and options are shown in Fig.~\ref{fig:butler-cli}.
We have defined a pluggable architecture where packages other than the core \texttt{daf\_butler} package can register their own Butler subcommands.
This allows, for example, specialist commands such as \texttt{ingest-photodiodes} and \texttt{make-discrete-skymap} to appear in the subcommand listing even though they are not at all generic functionality.

\subsection{Data Ingest}

A Butler repository is not very useful without containing any datasets.
Facilities are provided for ingesting externally-generated files, and for the generic ingesting tooling\footnote{\texttt{butler ingest-files}.} care must be taken to define the correct data coordinates for each dataset and to ensure that those dimension values have already been defined.
The \texttt{obs\_base} package\footnote{\url{https://github.com/lsst/obs_base}} provides a command specifically targeted at ingesting raw imaging data.
The software scans a directory tree for files matching the provided glob, uses the \texttt{astro\_metadata\_translator}\footnote{\url{https://astro-metadata-translator.lsst.io}} infrastructure to read the headers and translate the contents to a standard form, creates \texttt{exposure} records as needed, and then ingests the files by constructing a data coordinate from the translated metadata.
It can handle raw files that store all the detector images in a single file or as one file per detector, and Butler does allow a single file artifact to be associated with multiple datasets in the Registry.
In situations where the files may be ingested into Butler repositories multiple times, a facility is also provided to use ``sidecar'' JSON files containing extracted metadata, or even a JSON index file containing metadata for multiple files, since it much quicker to parse JSON than to extract and parse a FITS header, especially if that FITS file is in an object store.

\subsection{Calibrations}

As noted above, a \texttt{CALIBRATION} collection is a special type of collection that associates datasets with a validity range.
When a calibration is requested the data coordinate of the dataset being calibrated is used and any time-based dimensions (such as \texttt{exposure}) specify which calibration should be chosen.

Not all calibrations are calculated from datasets available to a Butler.
For example, for LSSTCam the QE curves are calculated by the camera team and given to the pipelines team as part of the camera delivery.
Additionally, some calibrations, such as defect masks, are relatively static, and these can be represented in a compact text format, and are useful to people outside of a Butler.
We call such datasets ``curated calibrations'' and store them in Git repositories using text file formats and a directory layout that makes the validity ranges clear.
It can be very convenient when setting up a new Butler for an instrument to be able to seed it with these curated calibrations without having to locate raw data files and reconstruct them from scratch, and infrastructure is provided to enable that, for example using the \texttt{butler write-curated-calibrations} command line tool.
Generally, when these curated calibrations are stored the Butler converts them from text format to a binary format to make reading them more efficient; this is all handled by the formatter configured for the particular Datastore.

\section{The Pipeline System}

\setcounter{footnote}{0}  % Latex only allows 9 footnotes with symbols.

Processing data is done with reusable units of code called \texttt{Task}s.
Each \texttt{Task} has a specialized configuration object attached to it (from the \texttt{pex\_config} package\footnote{\url{https://github.com/lsst/pex_config}}) and must provide a \texttt{run()} method that is the method that implements the algorithm.
\texttt{Task}s can be nested within each other in a hierarchy to create high level algorithms.
However, the top level \texttt{Task} of any algorithm must satisfy an additional interface, defined as a \texttt{PipelineTask}, which allows it to interact with the \texttt{Butler} for i/o and ordering within a processing pipeline.

A \texttt{PipelineTask} is special because it requires the author to declare the task's ``connections'': the datasets the task will consume as inputs and produce as outputs.
The interface also requires the definition of the task's dimensions, which defines the unit of processing over which the task runs.
The task's dimensions do not need to match those of its input or output datasets; for example, a task with dimensions \texttt{\{instrument, exposure\}} that takes an input with dimensions \texttt{\{instrument, exposure, detector\}} is a ``gather'' step that processes all detectors from that exposure together, but processes each exposure independently (and hence possibly in parallel).
Other examples of \texttt{PipelineTask} dimensions are \texttt{detector} and \texttt{exposure} which operates on data from an exposure of a single detector taken from the telescope, whilst another could be \texttt{band}, \texttt{tract}, and \texttt{patch} (sky regions) which would operate on many intermediate data products produced from previous processing stages that overlap the specified region.
The input and output datasets are tied in to the Butler by declaring the dataset type name and associated storage class.
The storage class is a proxy for the Python type and thus allows the pipeline author to guarantee that the correct Python type will be given (converting it from another type if required and supported) -- if the storage classes associated with datasets of that dataset type in the target Butler are not compatible with those required by the pipeline the pipeline will not run.

To support pipeline execution the \texttt{run()} method of a \texttt{PipelineTask} must take parameters that match the input connections and must return a \texttt{dict}-like data structure where the keys match the expected outputs.
A \texttt{PipelineTask} then has enough information to be able to pull the required datasets from a Butler and store the outputs, although task authors can override that default behavior using the associated config class to allow for more complexity in the Butler interaction.
Connections have options to allow the dataset loading to be deferred (something that is important when co-adding hundreds of images) or to allow multiple datasets to be provided if the dimensionality implies that.

\begin{figure}
  \lstset{language=YAML,basicstyle=\small\ttfamily}
  \begin{lstlisting}
    description: A demo pipeline.
    instrument: lsst.obs.subaru.HyperSuprimeCam
    tasks:
      calibrate:
        class: lsst.pipe.tasks.calibrate.CalibrateTask
        config:
          astrometry.matcher.maxOffsetPix: 300
      characterizeImage: lsst.pipe.tasks.characterizeImage.CharacterizeImageTask
      isr:
        class: lsst.ip.isr.IsrTask
        config:
          doVignette: true
          vignetteValue: 0.0
    \end{lstlisting}
  \caption{An example minimalist HSC pipeline describing single frame processing.}
  \label{fig:pipeline}
\end{figure}

Each \texttt{PipelineTask} provides one data processing step, several of which may be integrated together into a processing pipeline.
We define pipelines using a YAML text file; an example is shown in Fig.~\ref{fig:pipeline}.
The example pipeline consists of three tasks, with labels \texttt{calibrate}, \texttt{characterizeImage}, and \texttt{isr}, referencing the corresponding \texttt{PipelineTask} Python classes.
The order listed in the YAML file is not important because the classes implementing each task know which dataset types they need and the pipeline builder arranges them such that the output connections of one task are associated with the corresponding input connections of one or more other tasks.
Each task can have configurations specified in the pipeline file which override the default values, or overrides may be given through the pipeline executor command line interface.
This configuration not only supports algorithmic values, but also allows configuring the names used for the datasets in a task's connections.
The ability to configure dataset types provides flexibility to reuse the same \texttt{PipelineTask} with different configurations multiple times in a pipeline, with each configuration outputting a distinct dataset.
Additionally, it also provides flexibility in re-using pipelines.
Different pipelines can be derived from a common base pipeline, but tasks can be added or removed, and the datasets in the task connections can be configured such all the tasks can still be joined into a connected pipeline graph.
When a pipeline specifies an instrument explicitly, the \texttt{Instrument} class can be used to apply specific configuration overrides to any of the tasks in the pipeline.
Pipeline definitions can be significantly more complex than in the example.
To simplify this process, pipelines may be built up using the common recipe and ingredients paradigm\cite{2015A&C.....9...40J,2020ASPC..522..583L} to allow for pipeline reuse.
It is also possible for a user to specify that a subset of a pipeline be executed rather than the entire thing.

\begin{figure}
  (a) \includegraphics[width=0.95\textwidth]{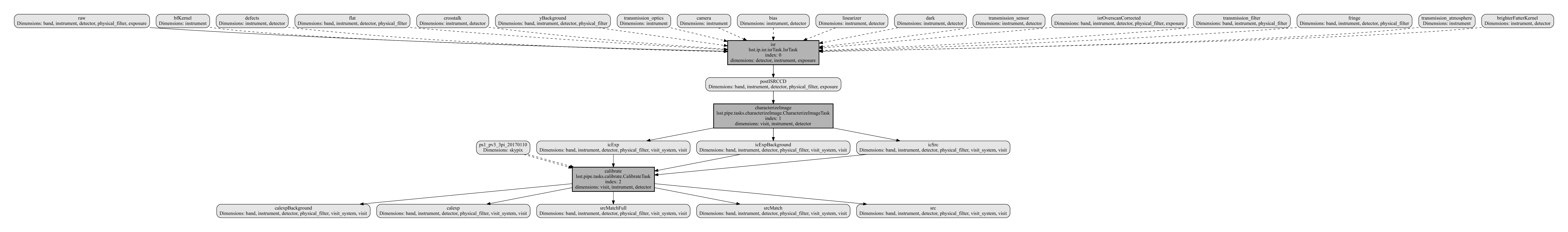}
  (b) \includegraphics[width=0.95\textwidth]{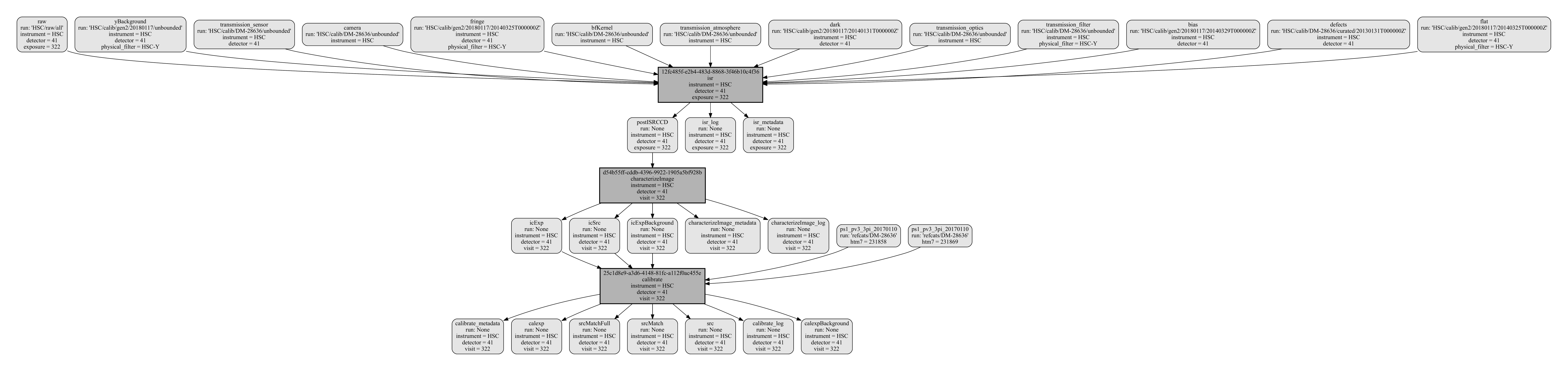}
  \caption{Top is a visualization of the pipeline shown in Fig.~\ref{fig:pipeline}. Bottom is the Quantum Graph for a single frame processing of one detector from one HSC observation showing the same pipeline with specific datasets.}
  \label{fig:graph}
  \end{figure}

Fig.~\ref{fig:graph}a shows a visualization of the demonstration pipeline shown in Fig.~\ref{fig:pipeline}.
The pipeline graph shows all the input and output datasets (including calibrations) and the order of execution.
What it does not show is any specific datasets attached to each step.
To determine what is to be processed a graph building algorithm queries a Butler and allocates datasets to tasks.
All of the datasets which correspond to a unique set of task dimensions are bundled together into a unit of work that we call a ``quantum''.
The set of all quanta to be executed and the relations between them and their tasks are called a ``Quantum Graph.''
The user provides an input Butler collection to search for datasets, in addition to a Registry query expression that affects not just the input datasets but the dimensions of the tasks and output datasets as well.
Having the Registry's dimension table skeleton populated before the Quantum Graph is generated allows us to query the database for both input datasets and predict future output datasets in much the same way.
The graph builder then allocates each matching input dataset to a specific Quantum and determines all the expected datasets that will be created during pipeline processing.
During processing the outputs are written to a \texttt{RUN} collection along with additional datasets containing all the log messages written by a task, software package versions, and processing metadata (including execution times and CPU usage for \texttt{Task}s).
In normal usage we then create a new \texttt{CHAINED} collection that combines the input collections and the new output \texttt{RUN} collection to allow a single collection to be queried for all the input and output datasets in that processing.
This procedure produces chains of collections, as processing campaigns proceed, which represent the final data products.
The Quantum Graph itself provides provenance information and is tied to the datasets by preserving the predicted output dataset UUIDs in the Butler when they are stored during processing.

\section{Integration with Batch Systems}

A Quantum Graph describes the work that the pipeline has to do in order to process all the data, but it does not say how the work should be scheduled or any resources that are required.
To run a Quantum Graph it has to be converted to something that can be executed.
On a single compute node (including a laptop) the \texttt{ctrl\_mpexec} package \footnote{\url{https://github.com/lsst/ctrl_mpexec}} provides a multi-processing executor, commonly run via the \texttt{pipetask} command line, that can run every quantum in the correct order and track all the responses.

When a Quantum Graph consists of thousands of quanta, a single compute node is no longer sufficient.
At that scale we need to use workflow systems such as HTCondor,\cite{10.1002/cpe.938} Pegasus,\cite{10.1016/j.future.2014.10.008} Parsl,\cite{10.1145/3307681.3325400} or PanDA.\cite{10.1088/1742-6596/331/7/072024}
For these workflow systems the Quantum Graph must be translated into the form expected by the target system.
Multiple prototypes were developed in parallel\cite{2020arXiv201106044B,DMTN-123,DMTN-137} but it was soon realized that many of the concepts for translating a Quantum Graph to a specialist graph are common and we developed a new Batch Production System (\texttt{ctrl\_bps}\footnote{\url{https://github.com/ctrl_bps}}) that would provide a framework to allow the aforementioned workflow systems to be treated as plugin code.

The BPS system provides a facility for a Quantum Graph to be converted to a generic form supporting workflow system concepts, called the Workflow Graph, before being converted to the final form.
This intermediate graph includes submission configuration parameters such as those that specify special resource requirements (CPU or memory) required by specific tasks.
The BPS submission process also allows quanta to be clustered for efficiency.
Some quanta can be quick, executing in tens of seconds, and not all batch systems are designed to efficiently handle such small payloads.
Clustering allows for multiple quanta to be combined into a single execution job; this can be particularly efficient if the clustering is designed such that the output of the first quantum is the only input to the next quantum (which is a common scenario for the early stages of many pipelines when a single detector is being characterized).

In the initial experiments, each quantum would talk to the main Butler repository when retrieving data at the start of the quantum and then when writing the results out at the end.
It soon became clear that this approach was not tenable once thousands of jobs are running concurrently and they are all trying to ask the Registry where their specific datasets are located.
We fixed this problem by changing the way that batch jobs interact with a Butler.
Initially we created a standalone read-only SQLite Butler Registry containing the relevant information for the entire workflow.
That did fix the problem with the large numbers of simultaneous connections and updates.
However, as a SQLite file it had to be copied to each job.
For very large workflows the SQLite file became large enough that the copying caused its own problems on some systems and at minimum caused time delays.
A better solution was to use the Quantum Graph itself since the graph can be stored in a shared location and only the appropriate subset of the graph needs to be read by each job.
The Quantum Graph knows all the input datasets and all the expected output datasets; it is therefore possible for a Butler to be constructed that uses the Quantum Graph itself as the Registry whilst using the main Datastore.
To enable this it was necessary to transfer Datastore records for existing datasets into the Quantum Graph.
Running large workflows in this manner means that the main Registry is not involved at all during the bulk of the processing.
Once the main workflow completes, a final job is executed that checks the Datastore to see which of the expected files were produced, and then ensures that the associated Registry entries are transferred back to the main Registry.
This merge job runs even if the workflow fails, to ensure that all the datasets that were generated are known to the main Registry.
These changes significantly improved the scalability of the bulk processing.
Note though that in this mode we are still writing to the main Datastore file system or object store and not using any file management facilities of the workflow system.
This decision is not hard-coded into BPS though and is entirely controlled by configuration.
In the future, we may decide to have jobs read and write to a job-local file system Datastore (whilst matching all the formatter and file naming configuration) and have the workflow system move files between the main Datastore and the job's Datastore.

\section{Development Process}

\setcounter{footnote}{0}  % Latex only allows 9 footnotes with symbols.

All the code described in this paper is written in Python (requiring at least Python version 3.9\cite{2020ASPC..522..541J}) and is open-source using the BSD 3-clause license for some packages and GPLv3 license for others.
The code is available from GitHub at \url{https://github.com/lsst} and we follow the Rubin Observatory Data Management development process.\footnote{\url{developer.lsst.io}}\cite{2018SPIE10707E..09J}
We use the \texttt{black} \footnote{\url{https://black.readthedocs.io/}} tool to automatically format the code, along with \texttt{isort} \footnote{\url{https://pycqa.github.io/isort/}} to order the Python imports.

Much of the code uses Python type annotations which are verified using the Mypy package.\footnote{\url{https://mypy-lang.org}}
We had a fairly large amount of code written before we decided to use type annotations.
It is known to be difficult to add type annotations to an existing project\cite{10.1109/ASE51524.2021.9678947}, and that was our experience.
There are still parts of the system that lack annotations and in some places it is extremely difficult to add them.
In particular, the more flexible an API is the harder annotations become, and this does begin to drive API design decisions.
We have, though, found that annotations do help once they have been added, and in particular code refactoring is less dangerous.

Documentation is built using Sphinx, based on the \texttt{documenteer} tooling\footnote{\url{https://documenteer.lsst.io}} and is integrated into the LSST Science Pipelines documentation.\footnote{\url{https://pipelines.lsst.io}}

We use GitHub Actions in all the repositories to ensure that code is formatted correctly, the type annotations are correct, and that all the tests pass and documentation builds.
We also automatically publish all the packages to the Python Package Index\footnote{For example the Butler can be found at \url{https://pypi.org/project/lsst-daf-butler/}.} when a tag is added to a repository.
These packages do not require the entire LSST Science Pipelines software system to be installed.

\section{Conclusions}

In this paper we have described a flexible system for abstracting data access from pipeline algorithmic code and for constructing complex pipelines.
The Butler system allows scientists and pipelines to have no knowledge of file formats or data locations, and allows for pipelines to be built in a manner where the pipeline builder can determine what data will be processed and how the individual algorithmic tasks will be combined.
This system has been successfully demonstrated at scale with the reprocessing of the DESC DC2 data\cite{2021ApJS..253...31L} on the LSST Interim Data Facility at Google \cite{2021arXiv211115030O} as part of LSST Data Preview 0.2 \cite{RTN-001} during late 2021 and early 2022.

\acknowledgments

We thank Eli Rykoff and Kian-Tat Lim for their reviews of the manuscript.
We also thank Kian-Tat Lim for his advice and feedback during the development of this system.
This material or work is supported in part by the National Science Foundation through Cooperative Agreement AST-1258333, Cooperative Support Agreement AST-1202910, and Cooperative Support Agreement AST-1836783 managed by the Association of Universities for Research in Astronomy (AURA), and the Department of Energy under Contract No.\ DE-AC02-76SF00515 with the SLAC National Accelerator Laboratory managed by Stanford University.
Additional Rubin Observatory funding comes from private donations, grants to universities, and in-kind support from LSSTC Institutional Members.

\end{document}